\documentclass[twocolumn, preprintnumbers,amsmath,amssymb,amsfonts,amsthm]{revtex4}
\usepackage{amsthm}% Include figure files
\usepackage{dcolumn}% Align table columns on decimal point
\usepackage{bm}% bold math

\newcommand{\gal}{{\rm Gal}}
\newcommand{\aut}{{\rm Aut}}

\newtheorem{prop}{Proposition}[section]

\theoremstyle{remark}

\theoremstyle{plain}

\newtheoremstyle{note}% name
  {3pt}%      Space above
  {3pt}%      Space below
  {}%         Body font
  {}%         Indent amount (empty = no indent, \parindent = para indent)
  {\itshape}% Thm head font
  {:}%        Punctuation after thm head
  {.5em}%     Space after thm head: " " = normal interword space;
  {}%         Thm head spec (can be left empty, meaning `normal')

\theoremstyle{note}

\newtheoremstyle{citing}% name
  {3pt}%      Space above, empty = `usual value'
  {3pt}%      Space below
  {\itshape}% Body font
  {}%         Indent amount (empty = no indent, \parindent = para indent)
  {\bfseries}% Thm head font
  {.}%        Punctuation after thm head
  {.5em}%     Space after thm head: " " = normal interword space;
  {\thmnote{#3}}% Thm head spec

\theoremstyle{citing}
\newtheoremstyle{break}% name
  {9pt}%      Space above, empty = `usual value'
  {9pt}%      Space below
  {\itshape}% Body font
  {}%         Indent amount (empty = no indent, \parindent = para indent)
  {\bfseries}% Thm head font
  {.}%        Punctuation after thm head
  {\newline}% Space after thm head: \newline = linebreak
  {}%         Thm head spec

\theoremstyle{break}

\theoremstyle{exercise}

\swapnumbers
\theoremstyle{plain}

\let\lvert=|\let\rvert=|

\begin{document}

%\preprint{APS/123-QED}

\title{$p$-adic CCR: Galois group representations, cyclic dynamics, and zeta-functions}% Force line breaks with \\

\author{Andreas Martin Lisewski}
% \altaffiliation[Also at ]{Belgradstrasse 19, 80796 Munich, Germany}%Lines break automatically or can be forced with \\
 \email{martin.lisewski@soundate.com}
\affiliation{%
Belgradstrasse 19, 80796 Munich, Germany
}%


\begin{abstract}
We consider a model of cyclic time evolution for Kochubei's $p$-adic realization of the canonical commutation relations (CCR). Connections to Kubota-Leopoldt $p$-adic zeta-functions and to arithmetic quantum theories such as the Bost-Connes model are examined.     
\end{abstract}

%\keywords{Decoherence; Semigroup; Measurement Problem}%Use showkeys class option if keyword
\maketitle

\section{\label{Introduction}Introduction}

The physical interpretation of arithmetic quantum theories has gained attention in recent years. Briefly, these quantum theories are those which exhibit Dirichlet $L$-series as thermal partition functions in their statistical formulation. Grounded on mathematical work of Julia \cite{julia} and  Bost-Connes \cite{boco}, it was realized that there exist similarities to certain aspects of string theory at finite temperature, or to other statistical systems such as the two-dimensional Ising model. For example, the pole of the Riemann $\zeta$-function $\zeta(\beta)$ at $\beta = 1$ may be understood as an indicator of a phase transition at Hagedorn temperature  $\beta^{-1} = 1$ \cite{julia, spector}, and the functional equation for the $\zeta$-function may be conceived as a duality relation in analogy to the Kramers-Wannier relation for the two-dimensional Ising model \cite{julia}. However, artihmetic quantum theories have not been treated as truly physical theories so far, which in other words means that until now none of them has been identified as a realistic model of a physical system.  

The motivation of this note is the question whether there are $p$-adic quantum systems that share certain aspects of arithmetic quantum theories. In contrast to conventional quantum theory, $p$-adic quantum systems such as the one investigated here are realized on spaces over the field $Q_p$ of $p$-adic numbers ($p$-adic fields, which in addition are algebraically closed are not regarded here.). Our main result is that the probably most elementary $p$-adic quantum system, namely the quantum harmonic oscillator whose generators satisfy the canonical commutation relations (CCR), exhibits a structural similarity to arithmetic quantum theories. Based on Kochubei's $p$-adic construction of the CCR \cite{kochu}, we show that the Galois group $\gal(Q_p/Q)$ of the abelian extension $Q_p$ of the rational numbers $ Q$ naturally leads to a representation $\rho$  on a one-dimensional vector space containing the oscillator's ground state. This leads us to a proposition stating that there exists a group homoemorphism from the Galois group $\gal(Q_p/Q)$ to the group of automorphisms acting on the algebra of bounded linear operators over the Banach space of continuous functions on $Z_p$. From the perspective of $C^*$-dynamical systems our representation motivates us to to regard $\gal(Q_p/Q)$ as the one-paramter group of cyclic time evolution for the $p$-adic quantum system. Further, the one-dimensional representation  itself turns out to be directly related to an Artin $p$-adic $L$-function associated to $\rho$. In our case the latter is equal to the Kubota-Leopoldt $p$-adic $\zeta$-function. We close with comments on the physical implications of our results and indicate obvious connections to the Bost-Connes arithmetic quantum system .

\section{$p$-adic CCR} 

We follow the work of Kochubei \cite{kochu}. Let $p$ be a prime number, $Q_p$ the field of $p$-adic numbers, $Z_p$ the ring of $p$-adic integers. Denote $C(Z_p, Q_p)$ the Banach space of $Q_p$-valued continuous functions on $Z_p$  equipped with $\sup$-norm. For $n \geq 1$ the functions
\begin{equation}
P_n(x) = \frac{x (x-1) \cdots (x-n+1)}{n!}
\end{equation}
with $P_0(x) := 1$ form complete orthonormal system, i.e. the Mahler basis, of $C(Z_p, Q_p)$. Thus every function $f \in C(Z_p, Q_p)$ admits a unique expansion
\begin{equation}
f(x) = \sum_{n=0}^{\infty} c_n \, P_n(x) \,, \quad c_n \in Q_p
\end{equation}  
with $|c_n|_p \rightarrow 0$, and $\|f\|_p = \sup_n|c_n|_p$, and with $| \cdot |_p$ being the absolue value of a $p$-adic number. Consider now the following operators on $C(Z_p, Q_p)$:
\begin{eqnarray}
(a^+f)(x) &=& x f(x - 1)\\
\nonumber
(a^-f)(x) &=& f(x +1) - f(x)\,.
\end{eqnarray}
This construction is well defined as it directly follows from the definition of the $\|\cdot\|_p$ norm that $a^{\pm}$ are bounded operators. A direct calculation of the commutator gives the CCR
\begin{equation}
[a^-,\, a^+] = 1 \,.
\end{equation}
Moreover, we obtain
\begin{eqnarray}
\nonumber
a^-P_n &=& P_{n-1}\,,\quad n\geq 1\,, \quad a^-P_0 = 0\\
\nonumber
a^+P_n &=& (n+1) \, P_{n+1}\,, \quad n \geq 0
 \end{eqnarray}
thus $a^+, a^-$ are clearly analogues of the creation and annihilation operators. Setting $H = a^+a^-$ we obtain $HP_n=nP_n$, and $[H, a^\pm]=\pm a^\pm$. The operator $ H$ has a complete system of eigenvectors and its discrete spectrum coincoincides with the set of non-negative integers $Z_+$. The whole spectrum of $ H$, however, equals $Z_p$, that is the $|\cdot|_p$-closure of $Z_+$ in $Q_p$. The density of $Z_+$ in $Z_p$ implies that the kernel of $a^-$ consists of constant functions on $Z_p$. This fact may be used to show that $a^-, a^+$ form an irreducible representation of the CCR. A remarkable property is that every Banach space over $Q_p$ having an infinite countable orthonormal basis is isomorphic to $ C(Z_p, Q_p)$.

\section{Ground state structure and Galois group representations}
As ground state of our $p$-adic quantum mechanical system we denote any continuous function $\Omega \in C(Z_p, Q_p)$ with $a^-\Omega = 0$ and $\Omega|_{Z_p}=1$. If we enlarge the domain of ground states and consider continuous functions $\Omega:X \rightarrow Q_p \in C(X,Q_p)$, $Z_p \subset X$, then we may expect more than just the trivial case, i.e. a globally constant mapping, because now any locally constant function on $Z_p$ with $\Omega|_{Z_p}=1$ may be regarded as a ground state, too. It is then convenient to look at  the equivalence class $\Omega$ of such ground states defined upon the relation
\begin{equation*}
\Omega_1 \sim \Omega_2 \Leftrightarrow \Omega_1, \Omega_2 \in C(X, Q_p): \Omega_1|_{Z_p} = \Omega_2|_{Z_p} = 1 \,.
\end{equation*}

As $Q_p$ is a field there are $p-1$ primitive roots of unity $\mu^{p-1}= \{\zeta_1,\ldots, \zeta_{p-1}\}, \zeta_j \in Z_p$ for all $j \in [1,p-1]$, together being the only roots of $x^{p-1} - 1$ in $Q_p$. These roots are the image of the Teichm\"uller character $\omega$, $\omega:Z^*_p \rightarrow \mu^{p-1}$, with the representation property $\omega(\alpha)\omega(\beta) = \omega(\alpha\beta)$, $\alpha, \beta \in Z^*_p$. Since $Q_p$ is an extension field of the field of the rational numbers $Q$, one has the corresponding Galois group $\gal(Q_p/Q) \simeq Z_p^*$. Consider the one-dimensional $Q_p$-vector space $V_0 = \{c \,\Omega:c \in Q_p\}$. Then, for every fixed $\kappa_0 \in \{0,\ldots, p-2\}$ the Teichm\"uller character $\omega$ admits a faithful one-dimensional representation $\rho_{\kappa_0}: \gal(Q_p/Q) \rightarrow \aut(V_0)$ of the Galois group, viz.
\begin{equation}
\label{rep1}
\rho_{\kappa_0}(\alpha)\,\Omega = \omega(\alpha)^{\kappa_0}\, \Omega, \, \, \forall \alpha \in Z^*_{p}\,.
\end{equation}
We stress that $\rho_{\kappa_0}$ is a continuous representation, since $\omega$ extends to a (uniformly) continuous function $\omega \in C(Z_p, Q_p)$. Since $Z^*_p \simeq Z_{p-1}$, equation (\ref{rep1}) may also be written as
\begin{equation}
\label{rep2}
\rho_{\kappa_0}(\alpha)\,\Omega = \,\zeta^{\kappa_0 \,t(\alpha)}_{p-1} \Omega\,,
\end{equation}
for a suitably chosen $t(\alpha) \in \{0,\ldots, p-2\}$ and $\zeta_{p-1}$ is the primitive root of unity of degree $p-1$.

Let $B(C(Z_p, Q_p)) \equiv B$ be the algebra over $Q_p$ of bounded linear operators on $C(Z_p, Q_p)$, i.e. $B = \{A \in B: \|Af\|<\infty \,,\,\forall f\in C(Z_p, Q_p)\}$. Our next task is to show that the above representation $\rho$ naturally induces a representation $\rho':\gal(Q_p/Q) \rightarrow \aut(B)$. Every $A \in B$ may uniquely be represented as a set $ A'$ of ordered pairs: $A' = \{(f, Af):f \in C(Z_p, Q_p)\}$. Obviously, the collection of all such sets forms an algebra $B' \simeq B$. We introduce the {\it van der Put basis} $\{e_n:n \in N_0\}$ of $C(Z_p, Q_p)$ as follows: $e_0 = 1$ and for $n >0$, $e_n$ is the characteristic function of the disc $D_n = \{x \in Z_p: |x - n|_p < 1/n\}$. Then for every $f \in C(Z_p, Q_p)$ one obtains the uniformly convergent series:
\begin{equation*}
f(x) = \sum_{n=0}^{\infty} v_n \, e_n(x) 
\end{equation*} 
with $v_0 = f(0)$ and $v_n=f(n) - f(n_-)$; $n_-$ is defined through the Hensel expansion given for any $n \in N$: $n = n_0 + n_1 p+ \ldots + n_sp^s$ with $n_s \neq 0$. Then $n_- = n_0 + n_1 p+ \ldots + n_{s-1}p^{s-1}$. Thus for every $A' \in B'$ and every $ f \in C(Z_p, Q_p)$ we have
\begin{equation}
(f, Af) = (\sum_{n=0}^{\infty} v_n(f) \, e_n(x), \sum_{n=0}^{\infty} v_n(Af) \, e_n(x)). 
\end{equation}
But every $e_n$ is locally constant on $ Z_p$; this is because $N$ is dense in $Z_p$ and each disc $D_n$ exhausts all the elements of $Z_p$. Hence, we may write\begin{equation}
(f, Af) = (\sum_{n=0}^{\infty} v_n(f) \, \Omega, \sum_{n=0}^{\infty} v_n(Af) \, \Omega)\,.
\end{equation} 
As the expansion in the van der Put basis is uniformly convergent, we have the follwing mapping due to equation (\ref{rep2})
\begin{equation}
\rho'_{\kappa_0}: t \mapsto (\sum_{n=0}^{\infty} v_n(f) \,\Omega, \sum_{n=0}^{\infty} v_n(Af) \, \zeta_{p-1}^{\kappa_0 t}\,\Omega)\,
\end{equation} 
with $t \in Z_{p-1}$. Since $\rho_{\kappa_0}$ is faithful this mapping determines an automorphism on $B'$, and so it does on $B$. Thus in summary, we have the proposition
\begin{prop}
\label{propo}
Let the tuple $(C(Z_p,Q_p), a^{\pm}, B)$ be the $p$-adic quantum mechanical system as previously defined. Let further $\Omega \in V_0$ denote the ground state of this system. Then for every $\kappa_0 \in Z_{p-1}$ there is a faithful representation $\rho'_{\kappa_0}:\gal(Q_p/ Q) \rightarrow \aut(B)$ uniquely determined by the one-dimensional continuous representation $\rho_{\kappa_0}:\gal(Q_p/ Q) \rightarrow\aut(V_0)$.
\end{prop}
\section{$p$-adic $\zeta$-functions}
Finite dimensional representations of Galois groups associated to field extensions exhibit a remarkable relation to certain $L$-functions. In particular, due to the pioneering work of Delgine and Ribet \cite{delrib} one can naturally define a $p$-adic $L$-function $L_p(s, \rho)$ to any even representation $\rho$ on a vector space over a $p$-adic field such as is $Q_p$. This function is referred to as $p$-adic Artin $L$-function associated to $\rho$. Let $C_p$ be the smallest extension field of $Q_p$ that is algebraically closed and complete with respect to $|\cdot|_p$. Then a one-dimensional representation of $\gal(C_p/Q)$ generally reads as 
\begin{equation*}
\rho(\sigma_\alpha) \, e = \chi(\alpha) \, e
\end{equation*} 
where $e$ is the basis of vector space $ V = C_p\, e$. Here, $\rho$ is even  when $\chi(-1) = 1$. For even representations of this kind the $p$-adic Artin $L$-function associated to $\rho$ becomes the $p$-adic Dirichlet $L$-function originally intrduced by Kubota and Leopoldt \cite{kubota}. If we now restrict our view to extension field $Q_p$ then the one-dimensional representations of $\gal(Q_p/Q)$ are those given by equation (\ref{rep1}) and - equivalently - by equation (\ref{rep2}) with even characters $\chi = \omega^{\kappa_0}$. In this case the Kubota-Leopoldt $p$-adic Dirichlet $L$-function
 is defined through a $Q_p$-valued measure $\mu(x)$ and it reads (For an introduction to $p$-adic inegration and $p$-adic $L$-functions, see \cite{koblitz}.)
\begin{equation}
\zeta_{p, \kappa_0}(s) = \frac{1}{\langle r \rangle^{1 - s} \omega(r)^{\kappa_0} - 1} \int\limits_{Z^*_{p}} \langle x\rangle^{-s} \omega(x)^{\kappa_0-1} \,d\mu(x)
\end{equation}  
where $r$ is any integer prime to $p$,  $\langle r \rangle = r/\omega$, $ \langle x\rangle = x/\omega$, and the parameter $\kappa_0 \in \{0,\ldots,p-2\}$ depicts the {\it branches} of the $p$-adic $\zeta$-function. $p$-adic Dirichlet $L$-functions have the remarkable property that they interpolate complex Dirichlet $L$-functions at algebraic values, i.e. values for $1-k$ with $k \in N_0$. So, for example, if we choose $p=2$ then we have
\begin{equation*}
\zeta_{2, 0}(1-k) = (1 - 2^{k-1}) \,\zeta(1-k) = (2^{k-1} - 1) \frac{B_k}{k}\,, 
\end{equation*}
where $\zeta$ is the Riemann $\zeta$-function and $B_k$ is the $k$th Bernoulli number. Despite this close analogy to complex Dirichlet $L$-functions, $p$-adic Dirichlet $L$-functions are less understood than their complex counterparts. For instance, a general functional equation for $p$-adic Dirichlet $L$-functions is unknown, and their values for natural numbers larger than one are not known either.  
\section{Discussion}
Proposition \ref{propo} plays a similar role for our $p$-adic quantum mechanical system as does a proper group homeomorphism $\alpha: R \rightarrow\aut(\cal{A})$ play for some $C^*$-dynamical system within the framework of conventional quantum mechanics, i.e. the role of time evolution within the algebra of observables.  In the latter, $\cal A$ usually is the $C^*$-algebra of bounded linear operators on some complex Hilbert space. Despite this obvious similarity there are significant differences. Recall that $Q_p$ is not an algebraically closed field; for example, the equation $x^2 + 1 = 0$ does not have a solution in $Q_p$. Therefore, our construction shows that with $p$-adic analysis over $Q_p$ it is possible to represent a non-trivial quantum mechanical system and its time evolution on a real (in the field-theoretic sense) one-dimensional vector space-- something that is totally missing for one-dimensional (Hilbert) spaces over $R$. In fact, if we try to apply our results to the case of real numbers and consequently identify $Q_\infty = R$ then the corresponding Galois group becomes $\gal(R/Q) \simeq \{{\rm id}\}$, i.e. the trivial group. Now the only roots of unity in $R$ are $\{\pm 1\}$. Since there is no continuous homeomorphism $\rho: \{{\rm id}\} \rightarrow \{\pm 1\}$, there is no non-trivial time evolution either.
For the $p$-adic case, on the other hand, we have a cyclic time structure defined for each branch that is  depicted by a value of $\kappa_0$. Surely, the physical meaning of a $p$-adic time parameter $t \in Z_{p-1}$ is far from being clear, because the usual complete ordering of real numbers is lost here. The latter property, however, has always been considered as being essential for a physical time parameter. So, $p$-adic quantum systems such as the one investigated here may indicate that $R$ does not give the only possible parametrization of time. 

Finally, we want to briefly mention an obvious relation  between the $p$-adic quantum system as introduced in this note and the arithmetic quantum system of Bost and Connes. In both systems the Galois group of an abelian extension of $Q$ has been represented on the corresponding algebra of observables. And in both systems a Dirichlet $L$-function naturally occurs as a consequence. In the Bost-Connes system this $L$-function is the Riemann $\zeta$-function $\zeta(\beta)$ which at the same time turns out to be the partition function of the quantum statistical system at temperature $\beta^{-1}$. Therefore, one may wonder whether it is admissable to regard $\zeta_{p,\kappa_0}(s)$ as a $p$-adic partition function, and whether $s$ may become the inverse $p$-adic temperature in any reasonable sense. Further, it is interesting to observe that our $p$-adic quantum system possesses an extra symmetry not present in the Bost-Connes system: the one-paramter group representation of $\gal(Q_p/Q)$ appears twice as realized through the parameters $\kappa_0$ and $t$. Thus $\gal(Q_p/Q)$ plays a double-role here in the sense that it is a symmetry group of the quantum system as well as it is the system's dynamical group representing time.   

\vfill
\bibliography{amlisewski}
\end{document}